\documentclass[preprint,11pt,fullpage,doublespace,linenumbers]{article}
\usepackage{hyperref}
\usepackage{graphics}
\usepackage{color}
\usepackage{amssymb}
\usepackage{graphics, slashed}
\usepackage{amsmath}
\usepackage{slashed}
\usepackage{chngcntr}
\usepackage{geometry}                
\usepackage{graphicx}
\usepackage{amssymb}
\usepackage{epstopdf}
\usepackage[normalem]{ulem} 
\usepackage {ulem} 
\usepackage{colordvi}

\counterwithin{figure}{section}
\usepackage[normalem]{ulem} 
\usepackage {ulem} 
\usepackage{color}
\usepackage{colordvi}
\usepackage{hyperref}
\def\ee{\end{equation}}
\def\be{\begin{equation}}
\def\eea{\end{eqnarray}}
\def\bea{\begin{eqnarray}}

\def\BT{Bruno Touschek}

\begin{document}
%
\title{The path to high-energy electron-positron colliders: from Wider\o e's betatron to Touschek's AdA  and to LEP}
\author{Giulia Pancheri $^1$\thanks{pancheri@lnf.infn.it}  \ and Luisa Bonolis $^2$\thanks{lbonolis@mpiwg-berlin.mpg.de}  }
\maketitle
\noindent 
$^1$ Laboratori Nazionali di Frascati dell'INFN, Frascati, I00044, Italy\\
$^2$  Research Program History of the Max Planck Society, Max Planck Institute for the History of Science, Boltzmannstra\ss e 22 - 14195 Berlin, Germany
\section{Introduction}\label{:intro}
\begin{figure}[htb]
\begin{center}
\resizebox{0.7\textwidth}{!}{
\includegraphics{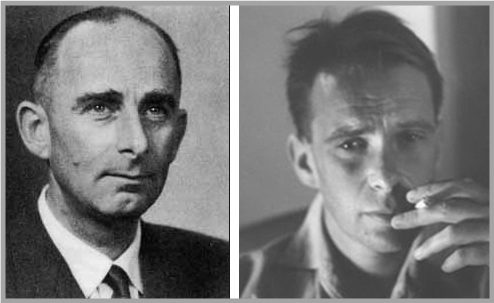}}
\caption{ Left:  Rolf Wider\o e, in middle age (\href{https://photos.aip.org/history-programs/niels-bohr-library/photos/wideroe-rolf-a1}{Emilio Segr\`e visual Archives, AIP}). Right: Bruno Touschek, 1955 (Courtesy of Touschek family).}
\label{fig:RWBT}
\end{center}
\end{figure}

In what follows, we describe the road which led to the construction and exploitation of electron positron colliders. We sketch the sequence of events which brought together the Austrian born Bruno Touschek \cite{Amaldi81}, shown in the right panel of Fig.~\ref{fig:RWBT} and the Norwegian Rolf Wider\o e \cite{Waloschek94}, shown in the left panel.
Their encounter, during WWII, in Germany, ultimately led to the construction in Frascati, Italy, in 1960, of the first electron-positron collider, the storage ring AdA, in Italian `Anello di Accumulazione', and to the first observation of  electron-positron collisions in a laboratory, in Orsay, France, in 1963.\\
Following this proof of feasibility, new particle colliders were then planned and built, leading to milestone discoveries such as the 1974 observation of the $J/\Psi$ particle, a new unstable state of matter, made of charm quarks. These early machines opened the way to the large particle accelerators of the 1980's, and the other discoveries that established the Standard Model of elementary particles as a well tested physics theory.

\section{The ante facts: in Europe before WWII}\label{sec:ante}
There are history making events, such as the construction of the first electron-positron collider AdA, which start by chance. In this section, we shall see how one of the last issues of the Physical Review to reach Nazi-occupied Norway in 1941, led Bruno Touschek  and Rolf Wider\o e   to meet and work together on the 15 MeV betatron project, commissioned by the Ministry of Aviation of the Third Reich, the Reichsluftfahrtministerium (RLM). This is how Touschek learnt from Wider\o e the art of making accelerators, enabling  him, many years later, to design, propose and carry through the construction of AdA.
\subsection{Bruno Touschek came from Vienna}\label{ss:vienna}
Bruno Touschek was born in Vienna, on February 3rd, 1921, and lived there until 1942, when he moved to Germany in order to pursue his studies in physics, which had been interrupted by the application of the Nuremberg Laws to Austrian citizens of Jewish origin. Bruno's maternal side was from a Jewish family prominent in the artistic circles of the Vienna Secession, the father a gentile and an officer in the Austrian Army.\footnote{Touschek's mother died young and his father later remarried.}
Because of his mixed parentage, Touschek's life and studies in Vienna were partly tolerated and he was allowed to attend the University of Vienna until June 1940, after which he was banned from classes and the University library.
Studying and learning became extremely difficult. Between June 1940 and Fall 1941 Bruno's physics education progressed thanks to Paul Urban (later Professor at University of Graz, Austria), who would borrow various textbooks from the University library and help him to study at his own home. Bruno thus was able to access Arnold Sommerfeld's fundamental treaty {\it Atombaum und Spektrallinien} summarizing all the development of the theory of atomic structure and of spectral lines, even finding small errors. Through Urban, Touschek had the chance to meet Sommerfeld in person in late 1941.\footnote{Edoardo Amaldi  \cite{Amaldi81}, in citing Urban, gives  the date November 24, 1942 for Urban's seminar.    While the month may be right, the year is actually 1941, as from letters exchanged between Urban, Sommerfeld, Touschek, and from Touschek's letters to his family in February 1942, already from Munich.}  Bruno's intelligence and drive in physics clearly impressed the great German theoretician. From the correspondence exchanged with Sommerfeld  and which followed this encounter there came the project for Touschek to move to Germany, where he could be able to unofficially attend classes and seminars held at University of Hamburg and Berlin by Sommerfeld's former students and friends, such as Max von Laue, Paul Harteck, Wilhelm Lenz.\footnote{The correspondence between Touschek and Sommerfeld is kept in the archives of the Deutsches Museum in Munich, Germany.}

In February 1942 Touschek left Vienna, first for Munich and then for Hamburg, where in August 1942 he was writing to his father, amid hopes for his future studies, mixed with homesickness, and money worries. He moved between Berlin and Hamburg, also visiting his family in Vienna occasionally, and attending unofficial seminars held there by Hans Thirring.\footnote{Touschek's life in Germany during the war is chronicled in a number of typewritten letters to his father and step-mother, to which GP and LB were given access by \BT's late wife, Mrs. Elspeth Touschek. These letters, including a few before the war, and a number of them extending well into the post-war period, {are presently in possession of their son Francis Touschek.}} To earn his living, he took small jobs, often obtained through Sommerfeld's friends, or by chance, as it happened when, in the train from Hamburg to Berlin, he met a girl, half-Jewish like him, who suggested he work for an electronic valve firm L\"owe-Opta in Berlin, furnishing electronic equipment to the war ministries, such as RLM. The firm was directed by Hans Thirring's former student, Karl A. Egerer, also editor and referee of scientific magazines such as the {\it Chemisches Zentralblatt}. In particular, and, most crucially, Egerer was the Editor-in-chief of the {\it Archiv f\"ur Elektrotechnik}, the journal, where, in 1928, Wider\o e had published his early article on the betatron \cite{RW28}. Bruno, who was charged with the establishment of a radio-frequency laboratory at L\"owe-Opta, became Egerer's editorial assistant and this is how, on February 10, 1943, he came across what, in a letter to his parents, he calls a ``sehr dummen Artikel'' [a very stupid article]. As we shall discuss below, the context of the letter, and the events which followed between February and June 1943, strongly suggest \cite{BP} that the article referred to was from the Norwegian engineer Rolf Wider\o e.
\subsection{Rolf Wider\o e was from Oslo}\label{ss:RW}
Rolf Wider\o e, the author of a proposal for a 15 MeV betatron (submitted to the {\it Archiv} on September 15, 1942) which reached Egerer and Touschek's desk in February 1943, was born in Oslo, July 1902. Following his parents' suggestion he went to study in Germany, first in Karlsruhe and then in Aachen, where he obtained his doctorate in engineering. Very early in his life, he had become fascinated \cite{Sorheim15} by the emerging field of particle accelerators. In 1928 he had described the successful construction of the first linear accelerator as part of his PhD thesis, based upon an earlier theoretical suggestion by Gustav Ising, and proposed a new technique to accelerate electrons and keep them in orbit, deriving what became known as the betatron equation \cite{RW28}. His linear accelerator (which may be considered as a `rolled out' cyclotron), and his first (unsuccessful) attempts to build a betatron, inspired Ernest Lawrence to build the first cyclotron and drew the attention of the American accelerator scientist Donald Kerst. Kerst succeeded in building the first betatron, and reported its operation in the {\it Physical Review} \cite{Kerst40-41}. 
These papers have been pivotal in the history of $e^+e^-$ colliders, firstly, as we shall describe below, because they played a crucial role in bringing Wider\o e back to Germany in 1943, secondly because, at around the same time, they became the subject of the graduation thesis of the Italian physicist Giorgio Salvini, the first director (1954-60) of the Frascati Laboratories.  Here, as we shall also see, an electron synchrotron of 1100 MeV went into operation in 1959, and, a few months later, the construction of the first electron-positron collider, AdA, was approved.

While in the United States advances in accelerator physics had been making history, Rolf Wider\o e, not having succeeded in building the betatron he was proposing, had left the active research field, and, in 1940, had joined the technical firm Norsk Elektrisk Brown Bovery in Oslo. However, he remained always interested in the field, and thus it happened that in late autumn 1941, he attended a Conference on particle accelerators, held in Oslo, at the Physics Association. Norway was occupied by Germany at the time and scientific communications with the English and American publications were dwindling, stopping altogether after the United States, in December 1941, entered the war. But one last copy of the Physical Review had reached the University of Trondheim by ordinary mail a few months earlier, with the article by D. Kerst on the first operational betatron.\footnote{Since the 1941 Physical Review articles were published in the July 1st 1941 issue, and reaching Norway by ordinary mail  would have taken 3 months, the journal   arrived not before October.}  Roald Tangen, Professor at the University, decided to make it the subject of a talk he was preparing for the Oslo Physics Association. When he gave his talk, he mentioned being intrigued by the reference to an article from an author with a Norwegian sounding name, R. Wider\o e \cite{Waloschek94}.\footnote{Notice that in Kerst's 1940 article, the reference to Wider\o e's paper \cite{RW28} has the wrong year. This was later corrected in the 1941 papers, but still with the wrong page number.}  Rolf was in the audience. Rushing back to his old passion, and clearly engaged anew seeing Kerst's success, in a few months Wider\o e prepared a proposal for a 15 MeV and even a 200 MeV betatron, and in September 1942 submitted it to the {\it Archiv}, where his original 1928 paper had appeared.
\subsection{Touschek and Wider\o e: the article which was never published and the 15 MeV betatron}\label{ss:RW-BY}
Between February and June 1943 there unfolded the events which led to the approval of Wider\o e's proposed project. Two of Touschek's letters to his family \cite{BP}, respectively dated February 15 and June 17, 1943, give a clue of what happened. From the first of such letters, one sees that Touschek discussed the physics of the project with his boss Karl Egerer. It also appears that he was critical of the treatment of relativistic effects, probably not correctly included at the time of the submitted article. Of special note, in this letter, is the mention of Heisenberg's possible interest,\footnote{At that time Werner Heisenberg 
had become the director of the Kaiser Wilhelm
Institute for Physics in Berlin 
and was leading the group
of physicists working on the German nuclear project, the  {\it Uranverein}. 
His auspicated interest provides an important premise towards the future involvement of the RLM in Wider\o e's proposal.} according to Egerer.

At the time, in Germany, there were other groups working on betatrons, such as Max Steenbeck at Siemens and the small company owned by Heinz Paul Schmellenmeier in Berlin. In Spring 1943, while the above mentioned machines were being developed, a project with a clear war potential was being considered by the Luftwaffe, a X-ray gun, proposed by Ernst Schiebold, a well known mineralogist and expert in X-ray technology. Wider\o e's higher energy betatrons, being a source of more powerful X rays, became connected with Schiebold's project as a possible realization of a death-ray weapon \cite{Waloschek2004}.

Wider\o e's article was brought to the attention of the military and soon taken into active consideration, to the extent that a few months later, March, or April, in Spring as Wider\o e says, he was approached in Oslo by ``several German Air Force officers'', who invited him to come to Berlin to discuss his project with the German military authorities. Wider\o e accepted and two days later \cite{Waloschek94} was flown to Berlin.
In the months that followed, Wider\o e refined his project, and a correspondence with Touschek started, about relativistic effects which needed to be taken into account. Touschek seems also to have joined the proposed project. In his letter of  June 17th, 1943,   Bruno Touschek writes:\\
``\dots slowly I begin even to be creditworthy. Based on my past work I was promised by the RLM, a monthly remuneration of about 100 marks. I have not seen anything yet \dots My Norwegian has been beaten on the whole line by me. Today there was a meeting in RLM, in which he had to agree with me point-by-point... A work in-print on the theory of a specific device will be pretty much changed based on my remarks."

Wider\o e worked on the article for the final publication expecting it to appear in the November issue  of the {\it Archiv f\"ur Elektrotechnik} (vol. 37, pp. 542-555), according to the reference mentioned in his autobiography. But the article, whose proofs are preserved in Wider\o e's archives in Zurich, is nowhere to be found in the 1943, nor in the 1944 issues of the journal. In fact it was never published, and a different article actually appears, exactly filling the range of pages corresponding to Wider\o e's article.\footnote{A preliminary copy of Wider\o e's article was sent to us by A. S\o rheim \cite{Sorheim15}. Copy of the article is actually preserved at the Archive of the Max Planck Society in Berlin, together with the undated typed draft of the second part describing a more powerful 200 MeV machine (Der Strahlentransformator II).}  Instead, in June, the project was approved by the German Aviation Ministry, due to start as early as possible, and classified as of military interest. In late summer (July or August?), Wider\o e went on vacation with his wife and had one of his inspirations: positive particles colliding against negative ones and releasing greater amount of energy than if hitting a stationary target. As Touschek later recalled about center of mass collisions: \\  ``The first suggestion to use crossed beams I have heard during war [was] from Wider\o e, the obvious reason for thinking about them being that one throws away a considerable amount of energy by using `sitting' targets'  most of the energy being wasted to pay for the motion of the center of mass."\footnote{Undated note, presumably written in February or March 1960.} 

Upon returning from his vacation Wider\o e mentioned his idea to Touschek and applied for a patent, dated as submitted on September 8th, 1943, but officially recorded only after the war, in 1953. Touschek scoffed at the idea, which he considered ``obvious", but the seed had been planted in his mind, to bear fruit when the time came, after the war and when both particle physics and the science of accelerators had progressed enough.

 The betatron project was housed near Hamburg, and in the months to follow, the project went ahead, being completed in late 1944. In March 1945, the war was at its end.   The military authorities ordered the betatron be moved to relative safety to a deserted factory house north of Hamburg, and as soon as this task was completed, on March 15 Touschek was arrested by the Gestapo and held prisoner in the infamous Fulsb\"uttel prison. But, as the Allied troops were approaching, orders came for the prisoners to be moved away, to the Kiel concentration camp 30 Kilometers North of Hamburg. As Touschek says in a postwar letter to his parents \cite{BP}:  ``There were SS soldiers behind us, in front, and on both sides."\\
He was lucky: as it is recalled in many slightly differing  versions,\footnote{Chronology and details of the events concerning Touschek's imprisonment and release are described by Touschek himself in two letters to his parents, dated June and October 1945 \cite{BP}, and do not coincide with Wider\o e's later recollections in \cite{Waloschek94}.}
he narrowly escaped death, and, after the war, went for his formal studies in physics first to G\"ottingen, for his diploma, and then Glasgow, for the DPhil. It is not clear why Touschek, who was part of Heisenberg's group, left and went to Scotland. Later Touschek regretted leaving G\"ottingen, but  in Glasgow a synchrotron was being built and Touschek was also involved in the project, thus furthering his knowledge on the art of making accelerators.  
After the doctorate and being awarded the Nuffield fellowship at University of Glasgow, Bruno was offered a position at University of Rome, and, in 1953, moved to Italy. 

In Norway, when the war ended in 1945, Wider\o e's decision to work for the occupying forces and the successful completion in Germany of the 15 MeV betatron by late 1944, were subjected to an inquiry by a Norwegian Commission, who accused Wider\o e of complicity with the occupying regime and even of participating to the construction of the V2 rockets which struck England during the war. Of this he was cleared, but his work in Germany was not condoned. Although he justified his collaboration with the Germans as a way to help reduce the pressure on his brother Viggo Wider\o e, a hero of the Norwegian resistance movement held prisoner in Germany at the time, he had to pay a heavy fine, which left him and his family in financial difficulties. Unlike Touschek, he did not turn to fundamental research, but instead went on to develop the field of radiation medicine in Europe, from Switzerland, where he moved in 1946 with his family, and at the Oslo Radium Hospital.

As for the betatron, it was appropriated as booty of war by the Allied Forces after the end of the war and brought to England, where it disappeared, after being used to X-ray iron plates \cite{Waloschek94}.
\section{AdA: from Frascati to Orsay}\label{sec:Frascati-Orsay}
After the war, the reconstruction of science in Europe developed alongside the creation of CERN and the decision to build particle accelerators as the tool for technological advances and for fundamental research in the laws of matter. 

The decision to create CERN and build a proton-synchrotron at the Geneva site went together with the establishment of national laboratories, equipped with accelerators, where scientists, technicians and engineers would be trained and become part of the national resources, on which CERN would draw. In France, it was decided to build  a proton synchrotron, SATURNE, and a linear accelerator (LINAC) of electrons, in Orsay, south of Paris. In Italy a circular machine, an electron synchrotron, was planned. The site was chosen to be near the town of Frascati, South-East of Rome. And so it happened that, at the end of 1959, in addition to SATURNE which had started operating in 1958, three powerful particle accelerators started functioning in Europe: at CERN it was the proton synchrotron, in Frascati the electron synchrotron, the LINAC in Orsay. With these accomplishments, new kinds of scientists, technicians and engineers, then called {\it machine builders}, prepared the field for the arrival of new accelerators, the particle colliders. In all the existing machines, particles,  protons or  electrons, were accelerated to hit stationary protons.  But, at this time, the concept of stacking beams of particles and making them collide head-on had begun to be investigated \cite{Kerst56,Oneill56}. 
Projects for
electron-electron storage rings were started in the US \cite{Oneill58},  as well as  in the Soviet Union under the direction of Gersh Budker,
first in
 Moskow and then  in the new laboratory of Novosibirsk in Siberia \cite{Skrinsky1996}. 

 In September 1959, a conference about the future of high-energy accelerators in physics was held at CERN. Both American and Russian scientist attended it.  Electron-positron collisions were also mentioned \cite{Panofski},
 but no one knew how to do it. 
As Touschek later  wrote: ``The challenge of course consist[ed]  in having the first machine in which particles which do not naturally live in the world which surrounds us  can be kept and conserved."\footnote{B. Touschek. Excerpts from a talk at Accademia dei Lincei, 24.5.74.} 
A few weeks after the CERN conference, Wolfgang K.H. Panofsky, from Stanford University, was in Rome giving a seminar   about the electron-electron rings being built in the USA, and  the question of using electrons and positrons was raised again, by Bruno Touschek.\footnote{The precise date of this seminar is not known, however a seminar by Panofsky at the Frascati National Laboratories on the subject of the Two mile linear accelerator in Stanford, is recorded as of October 27, 1959. It can be expected that the seminar in Rome would have taken place place around the same date.} Bruno Touschek clearly stressed that electron-positron annihilations -- reactions proceeding through a state of well-defined quantum numbers -- would be the  pathway to new physics.\footnote{N. Cabibbo in {\it Bruno Touschek and the Art of Physics}, docu-film by E. Agapito and L. Bonolis, INFN 2003. {Raul Gatto recalled that, ``Bruno kept insisting on CPT invariance, which would grant the same orbit for electrons and positrons inside the ring'' (personal communication to LB, January 15, 2004).}}
 
 \subsection{Frascati, Italy, 1960}\label{ss:Frascati}
Since arriving in Rome in 1953, Bruno had been engaged in theoretical work with colleagues at University of Rome, and had also become  actively interested in Frascati Laboratories, where the synchrotron had been built and, since December 1959, was now functioning. Thus, in the months following Panofsky's seminar, the idea of AdA started taking place and, on March 7th, 1960, during an epoch-making seminar at Frascati Laboratories, Bruno Touschek outlined his proposal to explore the physics of $e^+e^-$ annihilation processes and to build, as a first step, a small storage ring in which to store bunches of electrons and positrons to make them collide at a center of mass energy of 500 MeV.\footnote{Early notes by Touschek about the possibility of  electron-positron collisions dated as of 18.2.60,  follow a laboratory discussion about the future of the Frascati Laboratories on February 17th. Discussions about the physics to be explored  had also been taking place between Rome and Frascati, as seen by the paper by  N. Cabibbo and R. Gatto, ``Pion Form Factors from Possible High-Energy Electron-Positron Experiments" received by the {\it Physical Review Letters} on  February 17th, published on March 15.} {Before the seminar,  Touschek} had thoroughly explored the feasibility of what he was calling ``an experiment". In emphasizing the possibility of a complete transformation of the collision energy in the creation of new particles through a channel with the well-defined quantum numbers of the photon, Touschek saw in the inherent simplicity of the annihilation process, a conceptually different road to gain deep insight in fundamental particle physics and expressed his belief that this experiment should be ``the future goal" of Frascati Laboratories.

Only one week after the seminar, on March 14 the decision to build AdA was taken. Work to build AdA immediately started. Such was the optimism and the strong belief in this challenging enterprise that on November 9, 1960, Touschek prepared a note for a collider with an electron beam energy of 1500 MeV: \\
``\dots  if the work now in progress on ADA shows real promise (we will know more about this in February 1961) the development of ADONE should be considered as compulsory\dots ".\footnote{ADONE -- a Draft proposal for a colliding beam experiment, B. Touschek Archive, Sapienza University, Rome, Box 12, Folder 3.95.3.} \\
Indeed, on February 27, 1961, exactly one year after Touschek's first proposal,   the photomultiplier's current, which was recording the synchrotron light emitted by the circulating particles, showed that electrons were accumulated in AdA. In early June 1961, Touschek presented the first results at the International Conference on Theoretical Aspects of Very High-Energy Phenomena held at CERN in Geneva, announcing the development of two storage rings in Frascati, AdA and ADONE \cite{Cern1961}.
 
 \subsection{ AdA: Orsay 1961-64}\label{ss:orsay}
\begin{figure}[htb]
\begin{center}
\resizebox{0.8\textwidth}{!}{
\includegraphics{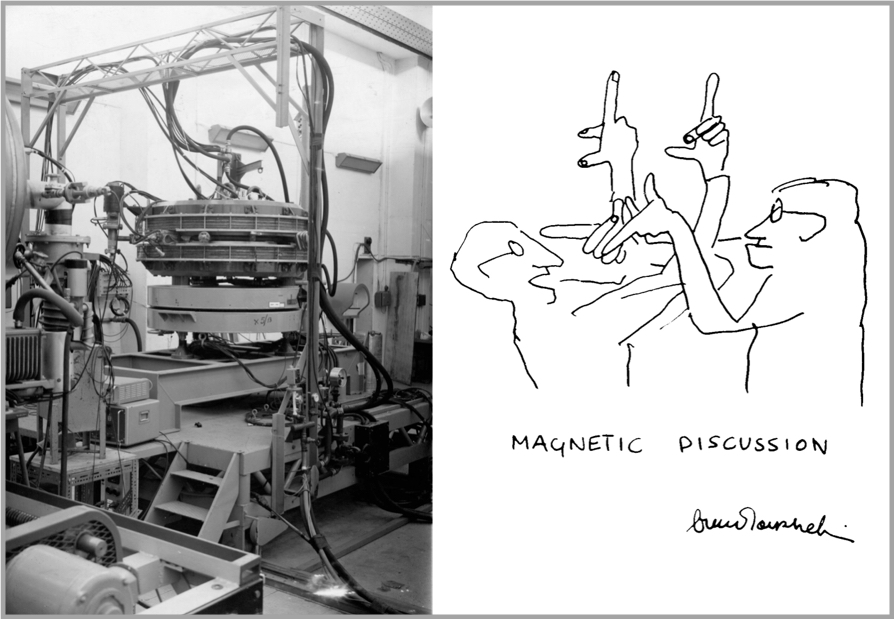}}
\caption{Left: AdA installed in Salle 500 MeV,  Laboratoire de l'Acc\'el\'erateur Lin\'eaire, Orsay, France, 1963. Right: a drawing by Bruno Touschek reflecting frequent  discussions about positrons or electrons entering AdA.}
\label{fig:ada}
\end{center}
\end{figure}

The announcement at the CERN conference had an immediate echo.\footnote{In addition to the French interest described in this section, we note the impact of Touschek's announcement  in  V.N. Baier, who    mentions it  as ``proof that we were not alone" in his reminiscences about {\it Forty years of Electron Positron Colliders}. The announcement is  also mentioned   in Baier's   early extensive paper on electron-positron physics, submitted in December 1962,  {\it Sov. Phys. Usp. 5 976}. } In the month of August, two French physicists, Pierre Marin from Orsay and Georges Charpak from CERN, went to Frascati to see the ``intriguing things'' happening there \cite{Marin}. Marin writes:\\
``In Frascati, a small team of top class physicists showed us with great pride a small machine, in a hall just next to the Frascati electronsynchrotron. This was AdA, {\it un vrai bijou}, a real jewel. The team, Bruno Touschek, Carlo Bernardini, Giorgio Ghigo, Gianfranco Corazza, Mario Puglisi, Ruggero Querzoli and Giuseppe di Giugno had succeeded in  storing in the ring  a few thousand electrons and positrons in turn, keeping them circulating for a few hours, until the beams decayed, observing them through the radiation they emitted. An enormous step for storage rings \dots".\\
However, the intensity of the gamma ray beam entering AdA    from the synchrotron   to produce  electrons and positrons within the ring was too low  and  the feasibility of the Frascati set-up for physics studies could not be demonstrated. It was then envisioned, among Marin, Touschek and Bernardini, that the machine could be moved to the Laboratoire de l'Acc\'el\'erateur Lin\'eaire, in Orsay, near Paris, in order to use the Orsay LINAC as a more efficient injector. In the year to follow between September 1961 and July 1962, an agreement between France and Italy was reached and on July 4th 1962 Ada was moved from Frascati to Orsay, where it was placed in the  500 MeV experimental hall (Salle 500 MeV), as shown in the left panel of Fig.~\ref{fig:ada}.

In France, thanks to the Orsay LINAC, the injection rate increased by no less than two orders of magnitude. Pierre Marin and Fran\c cois Lacoste from Orsay joined the Italian group. When later Lacoste left, Jacques Ha\"issinski joined the Franco-Italian team.\footnote{Ha\"issinki's 1965 Th\`ese d'\'Etat  is the most comprehensive  presentation of all that  was known at the time in electron storage ring physics.}  They had been given the use of the LINAC during the weekends, which meant that sometimes they would work as much as 60 hours in a row \cite{orsaymovie}. The turning point of  the experimentation in Orsay came one night, in late winter 1963. Carlo Bernardini remembers  \cite{CB}:\\
 ``The Linear accelerator was running like a dream, the injection of particles in the first beam worked like wonder (were they electrons or positrons?)." Touschek used to sketch their discussions in the drawing shown in Fig.~\ref{fig:ada}.\footnote{Bruno Touschek's drawings constitute an important chronicle of his life during the war, being included in his letters to his family, but also about academic life in Italy and, in particular about the student protests at University of Rome from  1968 onwards. Some of the drawings have been published in \cite{Amaldi81}, others in \cite{BP}. }\\
 And then something happened: the speed at which particles were being stored in the ring started decreasing, as if the beam lifetime were shortening.  The team panicked: was there a limit to the luminosity which could be achieved? Would they have to abandon the dream of higher energies, and forgo the future of this type of machines? Bruno decided to take a break, and left. It was about 3 a.m. in the morning. Then, on the paper covering a table at the Caf\`e de la Gare d'Orsay open all night, he sketched the solution of the problem, which he called the {\it AdA effect}. Touschek understood that there was a large momentum transfer from the radial to the longitudinal motion in a particle bunch due to the scattering of these particles within the same bunch. Many particles were lost when the density in the beam had reached a relevant value. This effect could have been devastating for the operation of the larger machine, such as Adone or ACO, the Anneau de Collisions d'Orsay which was being envisioned  by Pierre Marin, but Touschek had also calculated that this phenomenon was decreasing rapidly with energy. Still, due to the {\it AdA effect} (later the {\it Touschek effect})\cite{Touschekeffect}, AdA would suffer from an intra-beam scattering limit that gave a density-dependent lifetime generally significant at low energies (below 1 GeV, or so).

Although  the machine never reached the hoped-for collision rates for observing annihilation into new particles, AdA was however able to shed light on the actual size of the particle bunches, which had never been previously experimentally tested as far as the behaviour of stored beams was concerned. Most important, by means of single bremsstrahlung as a monitoring reaction, AdA was able to give the definite proof for the existence of conditions in which the two beams could be made to collide \cite{adaquartolavoro}.  The process $e^- +  e^+ \rightarrow  e^- + e^+ + \gamma$ was observed with AdA in Orsay, the interaction rate was measured and found to be in good agreement with the hypothesis that there was a complete overlap between the two beams and that the dimensions of the beams were those calculated from the lifetime effect.\footnote{The theoretical calculation for the cross-section of the single bremnsstrahlung process was discussed in the  {\it Tesi di Laurea} of Guido Altarelli and Franco Buccella, two of R. Gatt's students, published in the same issue of {\it Nuovo Cimento}  as the AdA paper of  
  Ref. \cite{adaquartolavoro}. } In  \cite{orsaymovie} Jacques Ha\"issinski says:\\
``C'\'etait la premi\`ere fois au monde que l'on montrait que les particules, effectivement, int\'eragissaient et entraient en collision les unes avec les autres et donc,  \c ca montrait que, on peut dire, que ces machines \'etaient utilisables pour faire de la physique des tr\`es hautes \'energies\dots".\footnote{``It was the first time in the world that it was shown that particles actually interacted and collided the one with the other and thus, this showed that, one could say, these machines could be used to make  physics at very high energies''.}

 \section{Electron-positron colliders: 1963- 1974}\label{sec:postada}
As 
 the words spread that it was possible to accumulate enough particles in the same ring and observe their collisions, projects for building higher energy accelerators based on these same principles, 
were 
{accelerated, }
 in the Soviet Union, in the United States, in Europe. 
 In Europe, France started building ACO, an electron positron collider with a center of mass (c.m.) energy of $2\times 550$ MeV, and Italy went ahead with the construction of a bigger and better AdA, namely ADONE with a c.m.  energy of 3000 MeV. 
 ADONE had been proposed by Touschek as early as November 1960, when he realized that AdA would soon be working. In the proposal to the Frascati Laboratory, there lies the decision which doomed Frascati not to observe the $J/\Psi$ at the same time as the Americans. Touschek's 1960 early proposal argued for a 3.0 GeV c.m. energy as being the highest energy to study pair production of all the elementary particles known to exists at the time, pions, kaons, including studies of the nucleon-antinucleon threshold around 2.0 GeV. In the Soviet Union   \cite{Skrinsky1996,Baier2006} the construction of VEPP-2, a 2000 MeV electron-positron collider, was accelerated in earnest, in the USA, at Stanford, an $e^+e^-$ ring with the highest energy of them all, the Stanford Positron Electron Accelerating Ring (SPEAR), with c.m. of 6.0 GeV, was approved with University funding, and built in a vacant lot owned by the University \cite{Paris2001}.  

VEPP-2 and ACO were the first to operate electron-positron storage rings with a high-energy physics program, where most of the basic storage ring machine physics was also studied and analyzed. Thus, in the late 1960's--early 1970s, a generation of powerful electron-positron colliders --- ADONE in Italy, ACO in France, SPEAR in California, DORIS at DESY and CESR at Cornell --- started operating. Expectations for new physics to explore and discover were high. In Frascati, in 1972 a surprisingly high rate of particle production was reported, a phenomenon of multihadronic production also observed at the Cambridge Electron Accelerator (CEA)  bypass, and  which signaled the appearance of new particles and possible application of the new theory of strong interactions, perturbative Quantum ChromoDynamics. In Stanford, as the c.m. energy rose to and passed the 3 GeV c.m. energy, the hadronic cross-section was observed to become anomalously large. Rumors in the US had it that a new state of matter was being created around this energy, a phenomenon observed also in hadronic fixed target collision. What became known as the November Revolution in particle physics started with a press conference in Stanford on November 11 1974, jointly held by Burton Richter, the director of the Stanford Linear Accelerator Center (SLAC), and Samuel Ting, the head of the Brookhaven experiment where the early signals of a bump in the cross-section had been observed. Frascati, with ADONE, confirmed the discovery on November 13th.\footnote{The news reached Frascati first by way of a telephone call from Sulan Wu, from the Brookhaven team, alerting Giorgio Bellettini, Director of the Laboratory at the time, and then from Mario Greco, staff member of the Frascati Laboratories, then at Stanford, who communicated  the exact value of the mass of the newly found particle.  This information was essential in order to tune the machine to  the right energy, since the $J/\Psi$ decays very rapidly, and  could be missed. Private communications by G. Bellettini and M. Greco.} The machine which had been invented by Touschek and had been built to discover new particles had succeeded in doing so, but it had come in second.

After the discovery of the $J/\Psi$, particle physics entered a new phase, something like a {\it Renaissance} period. Both proton\footnote{Proton colliders include the CERN  ISR and LHC (proton-proton),  a proton-antiproton  ISR run at 62 GeV, the CERN $S{\bar p}pS$, and FermiLab Tevatron (proton-antiproton),  and  HERA at DESY  (positron-proton).}
and new electron-positron colliders were planned and built, with a thousand-fold (and more) increase in the available center of mass energy, and accompanying luminosity. From AdA with 500 MeV c.m. energy and 4 meter circumference to the Large Electron Positron (LEP) collider with up to 209 GeV and 27 Kilometer around, from the Intersecting (proton-proton) Storage Ring at 53 GeV to the Large Hadron Collider in the LEP tunnel with up to 13 TeV c.m. proton-proton energy, particle colliders ushered in new discoveries and confirmed the validity of the Standard Model of particle physics. 

\section*{Acknowledgements}
We are grateful to Jacques Ha\"issinki for a  careful reading of the manuscript. We also acknowledge  helpful e-mail exchanges with S. Eidelman  about the Russian efforts.
G.P. thanks  the Institute of Physics for the opportunity to present this work at the Bristol Workshop on the History of Colliders, and the {\it European Physical Journal H} for supporting the presentation of parts of this material at the EPS History of Physics Conference in Pollau, Austria, September 2016. We also thank Touschek's family for access to the letters written by Bruno Touschek during the war.

\end{document}